\documentclass[runningheads]{llncs}
\usepackage[T1]{fontenc}
\usepackage{graphicx}
\usepackage{float}
\usepackage{cite}

\begin{document}
\title{Adversarial Stain Transfer to Study the Effect of Color Variation on Cell Instance Segmentation}
\titlerunning{Study the Effect of Color Variation on Cell Instance Segmentation}
\author{Huaqian Wu\inst{1} \and
Nicolas Souedet\inst{1} \and
Camille Mabillon\inst{1} \and
Caroline Jan\inst{1} \and \\
C\'edric Clouchoux\inst{2} \and
Thierry Delzescaux\inst{1}}
\authorrunning{H. Wu et al.}
\institute{CEA-CNRS-UMR 9199, MIRCen, Fontenay-aux-Roses, France
\email{thierry.delzescaux@cea.fr}\\
 \and
WITSEE, Paris, France}
\maketitle              % typeset the header of the contribution
\begin{abstract}
Stain color variation in histological images, caused by a variety of factors, is a challenge not only for the visual diagnosis of pathologists but also for cell segmentation algorithms. To eliminate the color variation, many stain normalization approaches have been proposed. However, most were designed for hematoxylin and eosin staining images and performed poorly on immunohistochemical staining images. Current cell segmentation methods systematically apply stain normalization as a preprocessing step, but the impact brought by color variation has not been quantitatively investigated yet. In this paper, we produced five groups of NeuN staining images with different colors. We applied a deep learning image-recoloring method to perform color transfer between histological image groups. Finally, we altered the color of a segmentation set and quantified the impact of color variation on cell segmentation. The results demonstrated the necessity of color normalization prior to subsequent analysis.

\keywords{Histological images \and Microscopy \and Stain transfer \and Generative Adversarial Network \and Cell Segmentation.}
\end{abstract}
\section{Introduction}
Cell segmentation is the first step of many biological applications in preclinical research. For example, neuron instance segmentation is the prerequisite for quantitative studies of neuron population, morphology and distribution to investigate brain aging and neurodegenerative diseases. Advances in Whole Slide Imaging (WSI) techniques allow scanning entire tissue sections at the cellular level, while processing such a massive amount of data is still challenging. To reduce the manual workload, many automatic segmentation approaches have been proposed. In particular, Deep Learning (DL) based methods have demonstrated higher accuracy and robustness than traditional segmentation methods~\cite{wu2022cross,liu2021panoptic}. Although DL methods are considered more robust, the inconsistency of stain color in the histological images exists as a critical issue. This problem is related to many factors in the staining procedure, such as dye type, solution concentration, staining duration, and room temperature. Even if the staining procedure is strictly standardized, variation may still occur during the digitization. The DL models often underperform when applied to images with colors different from the training dataset~\cite{tellez2019quantifying}. To this end, most DL-based segmentation methods systematically applied color normalization as a preprocessing step~\cite{cui2019deep,kumar2017dataset}. In contrast, the impact of the color variation on the segmentation is still not fully understood and inadequately quantified.

Many stain normalization methods have been proposed, which can be divided into two categories: conventional and deep learning-based. Histogram matching~\cite{coltuc2006exact} maps the histogram of the source image to that of the target image, treating color distribution independently of image content. The color transfer method~\cite{reinhard2001color} converts the image into \textit{l}\( \alpha \beta \) color space and matches mean and standard deviation of histograms. However, they perform badly when the color of the source and target images differ significantly. The fringing method~\cite{macenko2009method} and the structure-preserving color normalization~\cite{vahadane2016structure} normalize the stain vectors in the optical density (OD) space separately for each staining channel. They were developed specifically for color variation in hematoxylin-eosin (H\&E) staining histological images. Neurons are often stained with NeuN, which is a biomarker evidenced by using immunohistochemistry (IHC). Color normalization in IHC-stained histological images is a less explored research area. Recently, several DL-based approaches have been proposed. These approaches generally address the color normalization as a style transfer problem, using Generative Adversarial Networks (GANs)~\cite{shaban2019staingan, bentaieb2017adversarial} to learn the color distribution of the reference stain and apply it on the source image. Nevertheless, the trained model can only deal with the specific stain of the training set and it is not suitable for images with multiple stains. GANs were often used as "black boxes", in which we had no explicit control over the image details until the invention of SyleGAN~\cite{karras2019style}. By feeding the generator gradually with latent style vectors, StyleGAN can control the style and appearance of the generated image. Furthermore, they show that the color-related features are mainly determined in the fine layers (superior layers). Inspired by StyleGAN, Afifi \textit{et al}. proposed HistoGAN~\cite{afifi2021histogan}, which controls the color of GAN generated images by feeding the desired color histogram in the last two blocks of the generator. In particular, a variant of HistoGAN, named ReHistoGAN, is designed to generate a realistic image with the color of the target image and the content of the input source image.

In this paper, we created a dataset dedicated to exploring color transfer, containing five histological image groups of a mouse brain with controlled color variations. For the first time, we applied ReHistoGAN to recolor histological images while keeping the image content. The trained ReHistoGAN is not limited to a single stain target, which was the main drawback of previous GAN-based normalization methods. We recolored a cell segmentation test set of a macaque brain with the trained ReHistoGAN, using images of different colors as references, including five from the mouse set and one from the macaque set as a control group. To our knowledge, this is the first study to quantify the effect of color change on cell segmentation of the DL model. The results revealed that the segmentation degradation of the DL model was linked to the color difference

\begin{figure}[H]
\includegraphics[width=\textwidth]{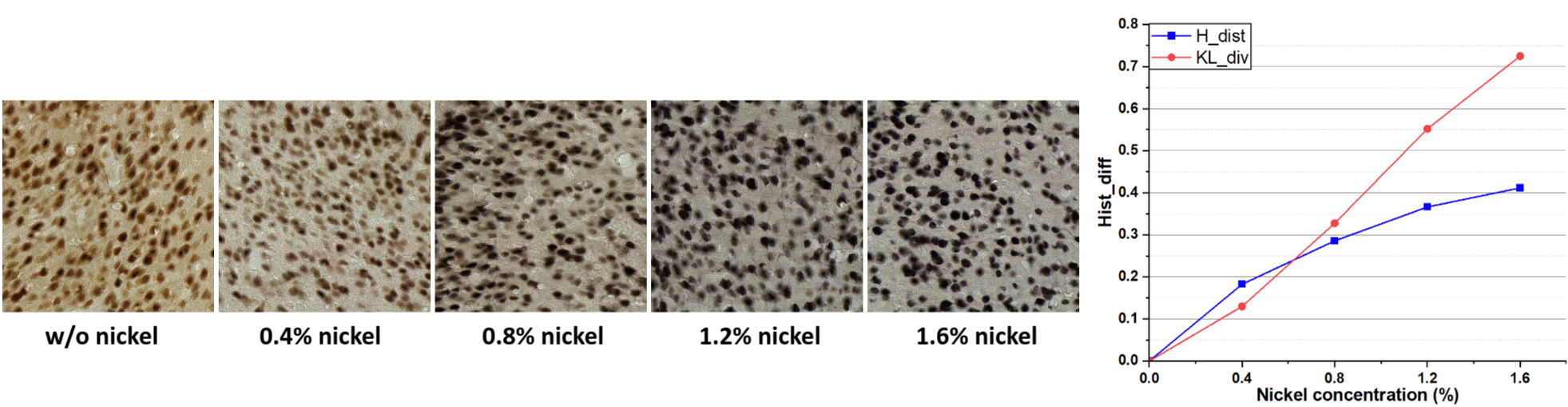}
\caption{Neuron microscopic images with five nickel concentrations. Left: image examples of different nickel groups. Right: correlation of histogram difference and nickel concentration (with the nickel-free group as reference).} \label{fig1}
\vspace{-5mm}
\end{figure}

\noindent between the training and test sets. The segmentation performance in groups with significant color variation was considerably reduced. On the contrary, the control group remains at the same level as the initial group. Thus, color normalization prior to inference is mandatory for robust and accurate segmentation.

\section{Material and Method}
\subsection{Dataset}
The data used for the color transfer study are mouse neuron microscopic images. Histological sections of a mouse brain were stained using IHC: sections were incubated with NeuN antibody, then treated with Diaminobenzidine (DAB) chromogen and nickel (a DAB enhancer) to develop color. DAB staining is usually brown, and the addition of nickel helps the staining become darker and easier to recognize. The DAB amount stayed constant throughout the staining process, and the only variable was the nickel concentration. This procedure yielded five section groups with different color representations, one of which was nickel-free and the other four were enhanced with nickel concentrations of 0.4\%, 0.8\%, 1.2\% and 1.6\%, respectively. The nickel concentration in this study is estimated by the proportion of nickel solution to DAB solution. Fig.~\ref{fig1} left illustrates image examples from the five groups. The higher the concentration of nickel, the darker the color of neurons. We manually extracted 10 images (3k $\times$ 3k pixels) of the cortex from each section group to build our dataset, including a training set of 13k 256 $\times$ 256 images and a test set of five 512 $\times$ 512 images, which represented five color groups. 

The cell segmentation dataset and a trained DL segmentation model were retrieved from~\cite{wu2022general}. Microscopic images were extracted from a macaque brain section stained with NeuN and nickel concentration of 1.6\%, knowing that the distribution and staining of neurons in mouse and macaque cortex are similar. This dataset was independent of the mouse color transfer dataset. In this study, we used only a subset of cortex, containing 36 1024 $\times$ 1024 images, on which we altered the stain color to examine the effect on segmentation.

\begin{figure}[H]
\includegraphics[width=\textwidth]{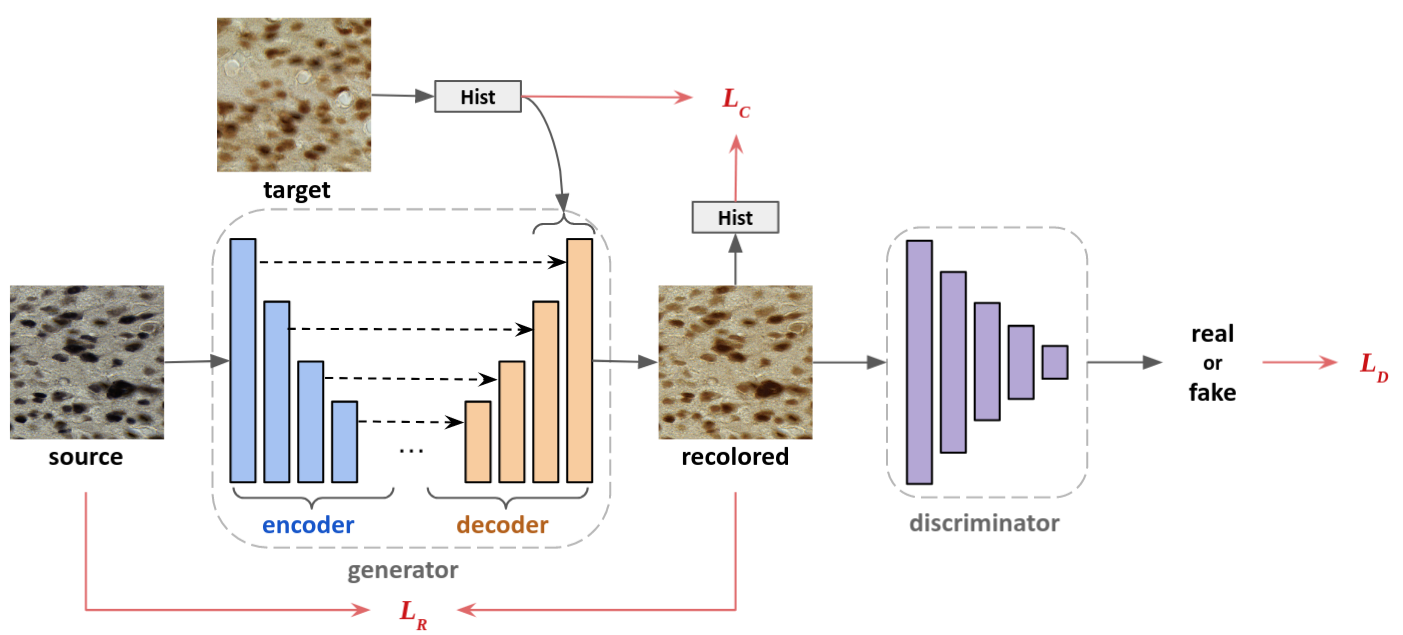}
\caption{ReHistoGAN architecture.} \label{fig2}
\vspace{-5mm}
\end{figure}

\subsection{Histogram feature}
We converted images from RGB to log-chroma space and calculated the histograms in this space. The intensity of each channel was normalized by the other two channels to make the histogram differentiable and insensitive to illumination changes. For example, the conversion of the R channel is defined as~\cite{afifi2021histogan}:
\begin{equation}
I_{uR}=log(\frac{I_{R}+\epsilon}{I_{G}+\epsilon}), I_{vR}=log(\frac{I_{R}+\epsilon}{I_{B}+\epsilon})
\end{equation}
where $I_K$ is the intensity of the $K$ channel of image $I$, $\epsilon$ is added for numeric stability, and the $u$ and $v$ are the $K$ channel normalized by the other two channels, respectively.
The $G$ and $B$ channels are also projected into log-chroma space to compute $I_uG$, $I_vG$, $I_uB$, $I_vB$. Binning is performed in u and v dimensions for each channel, resulting in a histogram with the shape of $n\times n\times 3$, where $n$ equals 64, is the number of bins.

The difference between the two histograms was measured using Hellinger distance (H dist) and Kullback-Leibler divergence (KL div)~\cite{afifi2021histogan}. They are usually used to quantify the similarity between two probability distributions. The more similar the distributions are, the closer the value is to 0. Previous studies suggested that H dist had a better sensibility to minor deviations, whereas KL div was more responsive to large deviations~\cite{afifi2021histogan, mocnik2021benford}. In this study, they were used to assess the color variation in different nickel groups and compare the color transfer performance of different normalization methods. In particular, the H dist was also employed to train the neural network as the color matching loss.

\subsection{ReHistoGAN}
The objective of ReHistoGAN was to generate a realistic image, preserving both the content of the source image and the color information of the target image. As shown in Fig.~\ref{fig2}, the network consisted of a generator to recolor the input image and a discriminator. The generator network was modified from a U-Net-like structure with skip connections. The target histogram features were projected into latent space and inserted in the last two blocks of the decoder to control the color of the output image. The generator was trained to generate an image having the same content as the source image and a color distribution similar to the target image. It is worth noting that the ReHistoGAN was trained in a fully unsupervised manner since the only ground truth needed was the histogram features, which were computed as described in 2.2. The goal of the discriminator was to distinguish the generated image from the real one. The loss function of the entire network is defined as~\cite{afifi2021histogan}:
\begin{equation}
L=\alpha L_{C}(H_{t}, H_{r})+\beta L_{R}(I_{s}, I_{r})+\gamma L_{D}
\end{equation}
where $t$, $s$ and $r$ are target, source and recolored images respectively. $H$ is the histogram. $L_C$ is the color matching loss computed with Hellinger distance. $I$ is perceptual detail (without color information), which was obtained with the Laplacian operator. $L_R$ is the reconstruction loss computed with the L1 norm. The discriminator loss $L_D$ is used to measure how realistic the generated image is. $\alpha $, $\beta $ and $\gamma $ are hyperparameters equal to 32, 1.5 and 4, respectively, which were defined empirically in~\cite{afifi2021histogan} to control the weight of each term. 

The ReHistoGAN model was then applied to recolor each image of the cell segmentation dataset using itself and five images in the mouse color transfer test set as targets. This procedure yielded six supplementary groups with continuous color changes. The first was a control group where no color changes were expected (same source image as target image), which was added to measure possible artifacts brought by the GAN network during encoding and decoding. The remaining five groups inherited the color distributions from the color transfer dataset, which were different from the segmentation set. Six colors on the same set allowed us to study the effect of color variation on cell segmentation. 

\subsection{Cell segmentation}
We utilized the segmentation method proposed in~\cite{wu2022general}, which consisted of a neural network to predict the cell, the inter-cell contour and the background, followed by a post-processing scheme based on mathematical morphology to individualize each cell instance. Color augmentation techniques were applied during the training to improve the robustness of the model to color inconsistencies. After training, we applied the network on the segmentation test set as well as the six GAN-recolored sets to study the impact of color variation. 

\subsection{Evaluation metrics}
In this study, the color variation between images was assessed by the difference between color histograms (H dist and KL div). A good color transfer method should provide a recolored image with low H dist and KL div matching the target. 

\begin{figure}[H]
\includegraphics[width=\textwidth]{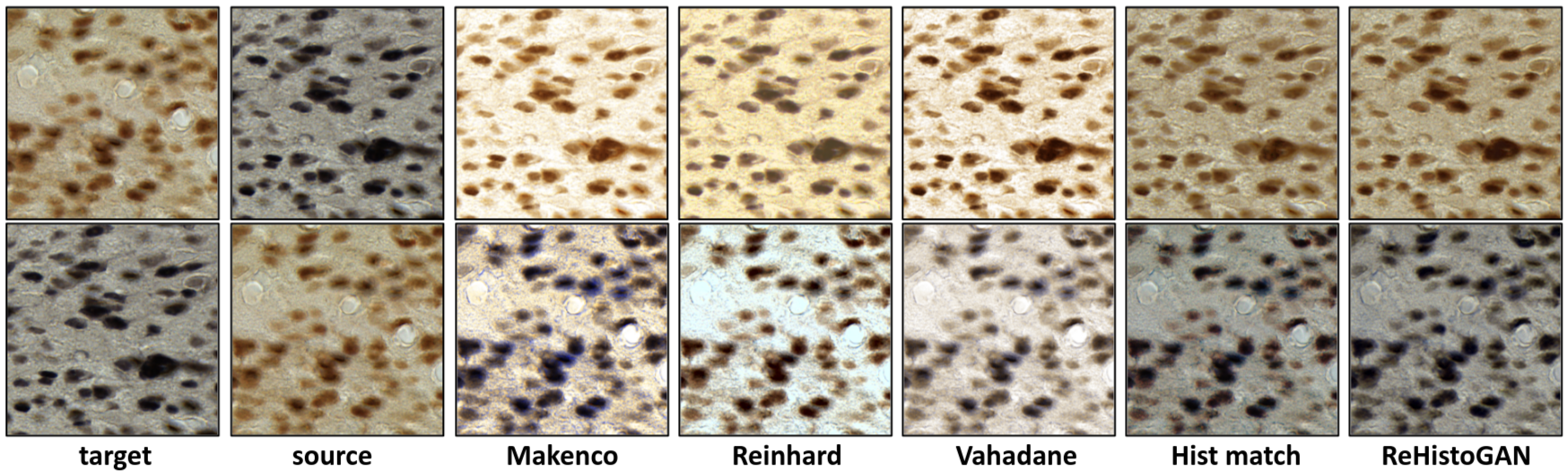}
\caption{Stain color transfer results from different methods. The first row presents recolored results of a 1.6\% nickel image using a nickel-free image as reference. The second row shows recolored results after switching the source and target.} \label{fig3}
\vspace{-10mm}
\end{figure}

\begin{table}[H]
\caption{Comparison of color transfer methods. The H dist and KL div are reported to show the color difference between recolored results and target images.}\label{tab1}
\begin{center}
\begin{tabular}{l|l|l|l|l|l}
\hline
Metrics & Macenko & Reinhard & Vahadane & Hist\ match & ReHistoGAN\\
\hline
H\_dist & 0.239\( \pm \)  0.038 & 0.284\( \pm \)  0.223 & 0.383 \( \pm \)  0.131 & 0.043\( \pm \)  0.031 & \textbf{0.052\( \pm \)  0.006}\\
KL\_div & 0.258\( \pm \)  0.076 & 0.42\( \pm \)  0.458 & 0.557 \( \pm \)  0.284 & 0.011\( \pm \)  0.012 & \textbf{0.011\( \pm \)  0.002}\\
\hline
\end{tabular}
\end{center}
\vspace{-10mm}
\end{table}

The cell segmentation was evaluated using three metrics: F1 score, Aggregated Jaccard Index (AJI) and Dice~\cite{wu2022general}. F1 score evaluated the cell instance segmentation, a segmented cell was considered a true positive when it overlapped the ground truth with an intersection over union greater than 0.5. AJI was a stricter metric to evaluate the instance segmentation since it penalized false segmentations. We also computed the Dice score to evaluate the semantic segmentation.

\section{Results and Discussion}
We measured the color histogram similarity between the nickel-free group versus other nickel groups using H dist and KL div. As shown in Fig.~\ref{fig1} right, color variation becomes more significant as nickel concentration gets higher. In the tested range, the KL div curve was roughly linear, whereas the H dist saturated as nickel concentration increased. 

\subsection{Color transfer performance of ReHistoGAN and other methods}
We compared ReHistoGAN with four state-of-the-art color transfer methods: Macenko \textit{et al}.~\cite{macenko2009method}, Reinhard \textit{et al}.~\cite{reinhard2001color}, Vahadane \textit{et al}.~\cite{vahadane2016structure} and histogram matching~\cite{coltuc2006exact}. Each image in the mouse test set was recolored using all color groups as color references, resulting in 25 combinations in total. Fig.~\ref{fig3} shows the visual comparison of two extreme cases from the nickel-free group (brown) and the highest-nickel group (dark). Macenko \textit{et al}.~\cite{macenko2009method} changed the stain color while it introduced artifacts, especially when recoloring the dark stain to brown. Reinhard \textit{et al}.~\cite{reinhard2001color} was unable to retain all color information of the target image, with the stain color in the results still affected by the source images. Vahadane \textit{et al}.~\cite{vahadane2016structure} achieved better results visually, however both Macenko \textit{et al}.~\cite{macenko2009method} and Vahadane \textit{et al}.~\cite{vahadane2016structure} suffered intensity issues, the recolored images were unnaturally too bright compared to the source and target images. The possible explanation for this might be that the IHC stain is a scatterer of light rather than an absorber. Therefore the Beer-Lambert law of absorption used for color space conversion is no longer applicable~\cite{vahadane2016structure}. The dark stain was successfully recolored into brown using the histogram matching~\cite{coltuc2006exact}. On the other hand, it did not perform well in converting brown stains to dark (brown spots still exist, see Fig.~\ref{fig3}), and the resulting stain colors were absent from both the source and target images. Compared to other methods, ReHistoGAN showed better performance and robustness, and it worked well in all cases, transferring the target color without losing the source content.

Table~\ref{tab1} reports the quantitative comparison of ReHistoGAN and other methods on mouse data. H dist and KL div of color histograms were measured between the recolored images and their target images. Histogram matching had the best average H dist and KL div since the objective was to produce similar color histograms. However, the high standard deviation suggests that this method lacks robustness. This finding is consistent with visual results in Fig.~\ref{fig3}, where the method performed poorly when the source and target images had very dissimilar statistics. On the other hand, besides comparable quantitative results as histogram matching (same KL div and slightly higher H dist), ReHistoGAN had the lowest standard deviations, which indicates its robustness.  Despite bringing illumination changes, Macenko \textit{et al}.~\cite{macenko2009method}, Reinhard \textit{et al}.~\cite{reinhard2001color} and Vahadane \textit{et al}.~\cite{vahadane2016structure} were unable to correctly adjust color variation in IHC stained images. Compared to Histogram matching~\cite{coltuc2006exact} and ReHistoGAN, these methods resulted in recolored images with more significant color differences from the target images.

\subsection{Impact of color variation on cell segmentation}
\begin{figure}[!t]
\includegraphics[width=\textwidth]{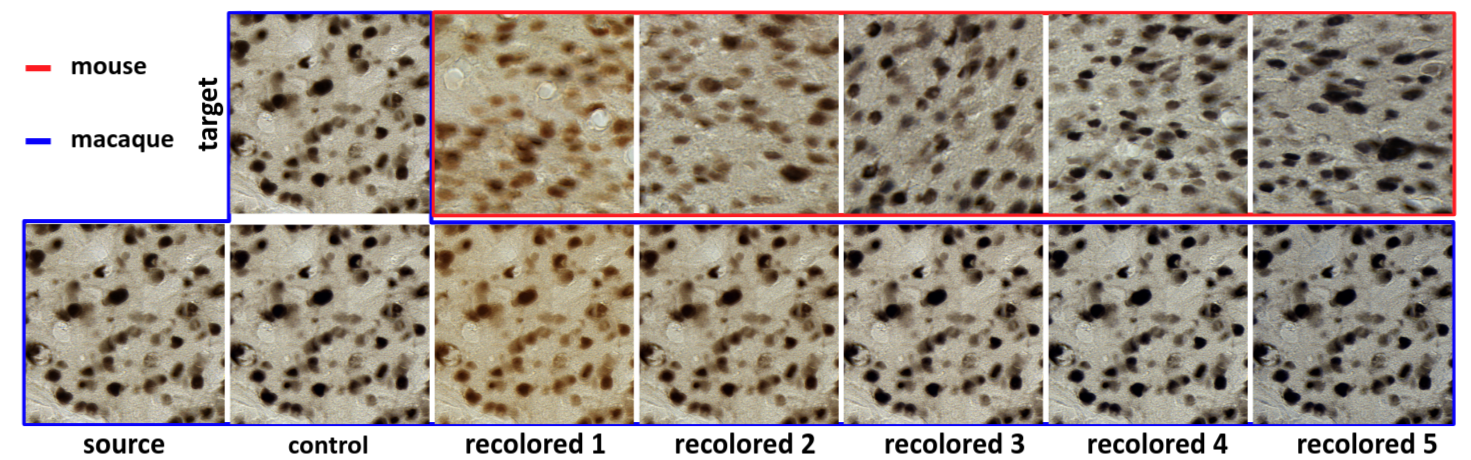}
\caption{Macaque segmentation dataset recolored by ReHistoGAN. First row: target images, second row: source macaque image and recolored results. From left to right: GAN-recolored images using the macaque image (control), nickel-free (recolored-1), 0.4\%, 0.8\%, 1.2\% and 1.6\% nickel mouse images (recolored-2-5) as targets.} \label{fig4}
\vspace{-5mm}
\end{figure}

We applied ReHistoGAN to recolor images of the macaque segmentation test set, as illustrated in Fig.~\ref{fig4}. The second column is the control group recolored using the same macaque image as the target, which allowed us to investigate the possible effect of encoding and decoding. The following columns show recolored images using five nickel mouse groups as the target, respectively. In total, six additional sets were produced, including one set without color change (control) and five with continuous color variations from brown to dark (recolored 1-5). Surprisingly, experts estimated that the original macaque images were closest in color to recolored-3 (0.8\%), although the nickel concentration used for macaque images was the same as recolor-5 (1.6\%). This inconsistency may be due to factors other than nickel concentration during the staining process.

\begin{table}[!t]
\caption{Comparison of segmentation performances (F1, AJI and Dice) on segmentation dataset without and with color variations inherited from the different groups in the color transfer dataset. Color variations between the two datasets are quantified using H dist and KL div.}\label{tab2}
\begin{tabular}{l|l|l|l|l|l|l|l}
\hline
Metrics & original & control & recolored-1 & recolored-2 & recolored-3 & recolored-4 & recolored-5\\
\hline
H\_dist & - & - & 0.257 & 0.093 & \textbf{0.05} & 0.132 & 0.182\\
KL\_div & - & - & 0.27 & 0.035 & \textbf{0.01} & 0.075 & 0.135\\
\hline
F1 & 0.91 & 0.899 & 0.842 & 0.892 & \textbf{0.898} & 0.893 & 0.867\\
AJI & 0.754 & 0.734 & 0.628 & 0.711 & \textbf{0.733} & 0.729 & 0.7\\
Dice & 0.972 & 0.961 & 0.909 & 0.951 & \textbf{0.962} & 0.956 & 0.936\\
\hline
\end{tabular}
\vspace{-10mm}
\end{table}

We evaluated the segmentation network on the original and six additional sets of macaque images. Table~\ref{tab2} reports the segmentation results and the color variation between the test sets and the training set. Despite the fact that the model was trained using color augmentation techniques, segmentation performance has been negatively impacted by color variations. The scores of the original set were the best since they came from the same data as the training set. They were used as references to show the performance degradation due to GAN artifacts and color variation. The scores of the control group were comparable to the reference but slightly lower, indicating that almost all information of the original image was preserved during encoding and decoding. The group with the most degradation in segmentation was the recolored-1 set, which had the most significant color variation compared to the original set (KL div: 0.27). In particular, the AJI of this set decreased by 17\% when compared to the reference. The recolored-3 set, on the other hand, scored similarly to the control group as it had similar color distributions both visually and quantitatively (KL div: 0.01). It suggests that the DL segmentation model is robust to slight color variation. Overall, the findings suggest that the segmentation performance decreases more in test sets with higher color variation. Surprisingly, the segmentation in the recolored-4 set was better than that in recolored-2, even though the former had a more visually distinct color variance. This might be due to the fact that despite the color variation, the contrast between cell and tissue is also essential for neural networks to segment cells. In our case, the darker stained images in the recolored-4 set presented a more significant contrast than lighter stained images in the recolored-2 set, providing a better condition for cell segmentation.

\section{Conclusion and Perspectives}
In this paper, we created an original dataset with five controlled color changes, which is well-suited for evaluating the performance of color transfer methods. We applied the ReHistoGAN method to perform the stain color transfer on IHC staining histological images. The results demonstrated its superiority and robustness compared with other state-of-the-art methods. Using ReHistoGAN, we intentionally altered the color of the macaque segmentation test set, which enabled us for the first time to quantitatively investigate the impact of color variation on IHC stained cell segmentation (semantic and instance) using neural networks. Experiments showed that the cell segmentation negatively correlated to the color variation between the test and the training sets. However, for a given color variation, the segmentation was better on images with darker stains. (see recolored-2 and recolored-4 in Table 2). Previous GAN-based approaches~\cite{bentaieb2017adversarial, shaban2019staingan} have not been assessed due to their lack of flexibility. Further studies are needed to validate the color transfer performance of ReHistoGAN compared to these methods. As a preliminary study, we only investigated NeuN staining images in the cortex region. Further work is needed to establish a strategy for the automatic selection of appropriate target images in various anatomical regions, which will allow us to expand this study to the entire brain. Moreover, additional study on images with other stainings (\textit{e.g.}, H\&E) would also be worthwhile.

\subsubsection{Acknowledgements.} This work was supported by DIM ELICIT grants from R\'egion Ile-de-France, by the French National Research Agency (project SUMMIT ANR-21-CE45-0022-01) and by the European Union's Horizon 2020 research and innovation program under the grant agreement No. 945539 (Human Brain Project SGA3).

\bibliographystyle{splncs04}
\bibliography{citations.bib}

\end{document}